\documentclass{aastex61}

\newcommand\aastex{AAS\TeX}

\received{}
\revised{}
\accepted{}
\submitjournal{ApJ}

\shorttitle{\aastex\ sample article}
\shortauthors{Zaqarashvili}
\begin{document}

\title{Equatorial magnetohydrodynamic shallow water waves in the solar tachocline}

\email{teimuraz.zaqarashvili@oeaw.ac.at}

\author{Teimuraz Zaqarashvili}
\affiliation{Space Research Institute, Austrian Academy of Sciences, Schmiedlstrasse 6, 8042 Graz, Austria}
\affiliation{Abastumani Astrophysical Observatory at Ilia State University, 0162 Tbilisi, Georgia}
\affiliation{Institute of Physics, IGAM, University of Graz, Universit\"atsplatz 5, 8010, Graz, Austria}

\begin{abstract}
The influence of a toroidal magnetic field on the dynamics of shallow water waves in the solar tachocline is studied. A sub-adiabatic temperature gradient in the upper overshoot layer of the tachocline causes significant reduction of surface gravity speed, which leads to trapping of the waves near the equator and to an increase of the Rossby wave period up to the timescale of solar cycles. Dispersion relations of all equatorial magnetohydrodynamic (MHD) shallow water waves are obtained in the upper tachocline conditions and solved analytically and numerically. It is found that the toroidal magnetic field splits equatorial Rossby and Rossby-gravity waves into fast and slow modes. For a reasonable value of reduced gravity, global equatorial fast magneto-Rossby waves (with the spatial scale of equatorial extent) have a periodicity of 11 years, matching the timescale of activity cycles. The solutions are confined around the equator between latitudes $\pm 20^0-40^0$, coinciding with sunspot activity belts. Equatorial slow magneto-Rossby waves have a periodicity of 90-100 yr, resembling the observed long-term modulation of cycle strength, i.e., the Gleissberg cycle. Equatorial magneto-Kelvin and slow magneto-Rossby-gravity waves have the periodicity of 1-2 years and may correspond to observed annual and quasi-biennial oscillations. Equatorial fast magneto-Rossby-gravity and magneto-inertia-gravity waves have periods of hundreds of days and might be responsible for observed Rieger-type periodicity. Consequently, the equatorial MHD shallow water waves in the upper overshoot tachocline may capture all timescales of observed variations in solar activity, but detailed analytical and numerical studies are necessary to make a firm conclusion toward the connection of the waves to the solar dynamo.
 
\end{abstract}


\section{Introduction} \label{sec:intro}

Rossby waves govern the large-scale dynamics of Earth's atmosphere and oceans \citep{Rossby1939,gill82}. They arise owing to the conservation of absolute vorticity and have been also considered to have an important influence on the dynamics of astrophysical discs \citep{Lovelace1978, Lovelace1999,Umurhan2010}, solar-like stars \citep{Lanza2009,Bonomo2012}, neutron stars \citep{Andersson1999,Lou2001,Heng2009}, planetary  \citep{Petviashvili1980} and exo-planetary atmospheres \citep{Heng2014}. Rossby waves have been used to explain the observed mid-range (or Rieger-type) periodicity in solar activity, which is seen in many activity indices such as sunspot number, radio flux, flares, and coronal mass ejections \citep{Rieger1984,carball90,Oliver1998,Lou2003}. Recent direct observation of Rossby waves in coronal bright points \citep{McIntosh2017}  has confirmed their important role in the large-scale dynamics of the solar atmosphere and interior.

The Coriolis parameter, which is actually the vorticity of a rotating sphere, depends on latitude with a minimum at the equator. Therefore, the equatorial region serves as a cavity for various shallow water waves (by analogy with the quantum harmonic oscillator, where potential walls trap oscillations). These so-called equatorially trapped or equatorial waves have been intensively studied by \citet{Matsuno1966}, \citet{Longuet-Higgins1968} and more recently by \citet{bouchut2005}. \citet{Lou2000} suggested that the equatorially trapped Rossby and inertia-gravity waves in the solar photosphere may explain the Rieger-type periodicity in solar activity. The consideration of \citet{Lou2000} is based on hydrodynamic (HD) shallow water equations, while the observed strong magnetic field on the Sun naturally supposes magnetohydrodynamic (MHD) effects.

\citet{Gilman2000} presented MHD shallow water equations for a horizontal magnetic field, which have been used for various applications to the solar tachocline \citep{Dikpati2001,Dikpati2003,Dikpati2006,Dikpati2007,zaqarashvili2010a,zaqarashvili2010b,zaqarashvili2015}. \citet{Schecter2001} studied various MHD shallow water waves in the f-plane approximation in the solar tachocline, but they did not consider equatorially trapped waves.
\citet{Zeitlin2013} studied shallow water MHD waves in $f$- and  $\beta$-planes using a quasi-geostrophic approximation and found that the magnetic field has a stabilising effect on the baroclinic instability.
\citet{Dikpati1999} and  \citet{Dikpati2003,Dikpati2006,Dikpati2007} performed a detailed stability analysis of a shallow water system under the action of the horizontal magnetic field with various latitudinal profiles and latitudinal differential rotation. It was found that the joint action of the differential rotation and the horizontal magnetic field generally leads to instability of the shallow water system. Nonlinear parametric interaction may also cause a mutual energy transformation between various waves in the MHD shallow water approximation \citep{Klimachkov2017}.

Despite significant work based on the MHD shallow water approximation, the linear spectrum of equatorial waves has not been studied in detail in the tachocline with a horizontal magnetic field. This consideration could be important for two main reasons. First, the magnetic field may significantly influence the spectrum of HD shallow water waves leading to, e.g., their splitting\citep{Lou1987,zaqarashvili2007,zaqarashvili2009,Heifetz2015}. Second, a sub-adiabatic temperature gradient in the upper overshoot part of the tachocline creates a negative buoyancy force, leading to reduced gravity \citep{Gilman2000,Dikpati2001}, which may cause a significant increase in Rossby wave timescales up to the periodicity of solar cycles.

Here we use the MHD shallow water equations to study the spectrum of equatorial waves in the solar tachocline.

\section{Governing equations}

We consider the solar tachocline as a shallow layer with thickness $H$ ($\sim 10^9$ cm) located at a distance $R$ ($\sim 5 \cdot 10^{10}$ cm) from the solar center \citep{Spiegel1992}.
We consider a local Cartesian frame $(x,y,z)$ on the rotating Sun, where $x$ is directed westward (i.e. in the direction of solar rotation), $y$ is directed northward, and $z$ is directed vertically outward. The layer  is traversed by an unperturbed toroidal magnetic field which is generally latitude-dependent: $B_x(y)$. We adopt solid body rotation with angular velocity - $\Omega$=2.6$\times$10$^{-6}$~ s$^{-1}$. Differential rotation is neglected at this stage for two reasons. First, the value of differential rotation is small near the equator. Second, it is important only for instabilities and will not significantly affect the wave dispersion relations.

We start with the linear shallow water MHD equations in a rotating Cartesian frame \citep{Gilman2000}
\begin{equation}\label{eq1}
{{\partial u_x}\over {\partial t}}-fu_y=-g{{\partial h}\over {\partial x}}+{B_x\over {4\pi\rho}}{{\partial b_x}\over {\partial x}}+{b_y\over {4\pi\rho}}{{\partial B_x}\over  {\partial y}},
\end{equation}
\begin{equation}\label{eq2}
{{\partial u_y}\over {\partial t}}+fu_x=-g{{\partial h}\over {\partial y}}+{B_x\over {4\pi\rho}}{{\partial b_y}\over {\partial x}},
\end{equation}
\begin{equation}\label{eq3}
{{\partial b_x}\over {\partial t}}+u_y{{\partial B_x}\over {\partial y}}=B_x{{\partial u_x}\over {\partial x}},
\end{equation}
\begin{equation}\label{eq4}
{{\partial b_y}\over {\partial t}}=B_x{{\partial u_y}\over {\partial x}},
\end{equation}
\begin{equation}\label{eq5}
{{\partial h}\over {\partial t}}+H{{\partial u_x}\over {\partial x}}+H{{\partial u_y}\over {\partial y}}=0,
\end{equation}
where $u_x$ and $u_y$ are the velocity perturbations, $b_x$ and $b_y$ are the magnetic field perturbations, $h$ is the perturbation of layer thickness, $g$ is the reduced gravity, $\rho$ is the fluid density and $f=2\Omega \sin \theta$ is the Coriolis parameter with $\theta$ being a latitude. Here we use Cartesian coordinates, which capture essential dynamics of large-scale equatorially trapped waves \citep{Matsuno1966, Lou2000}.  Cartesian coordinates enable an easier treatment of the wave dispersion relations, which are similar to those obtained under spherical geometry \citep{Matsuno1966,Longuet-Higgins1968}.   

Reduced gravity, which is an essential part of the shallow water system in the tachocline, is due to the sub-adiabatic temperature gradient providing a negative buoyancy force to the deformed upper surface \citep{Gilman2000}. Therefore, the surface feels less gravitational field compared to real gravity. The negative buoyancy force is proportional to the fractional difference between the actual and adiabatic temperature gradients $|\nabla-\nabla_{ad}|$, which is in the range of $10^{-4}-10^{-6}$ in the upper overshoot part of the tachocline and may reach up to $10^{-1}$ in the lower radiative part \citep{Dikpati2001}. Those authors showed that the dimensionless value of reduced gravity $G=gH/(R^2 \Omega^2)$ is proportional to $10^3 |\nabla-\nabla_{ad}|$, therefore it is in the range of $G>100$ in the radiative part of the tachocline and in the range of $10^{-3} \leq G \leq 10^{-1}$ in the upper overshoot part.

Differentiating Equations~(\ref{eq1}) and (\ref{eq2}) by time and inserting the time derivatives of $b_x$, $b_y$ and $h$ from Equations~(\ref{eq3})-(\ref{eq5}) we arrive at the equations
\begin{equation}\label{eq6}
\left [{{\partial^2 }\over {\partial t^2}}-(c^2+v^2_A){{\partial^2 }\over {\partial x^2}}\right ] u_x=\left [f{{\partial}\over {\partial t}}+c^2{{\partial^2 }\over {\partial x \partial y}}\right ] u_y,
\end{equation}
\begin{equation}\label{eq7}
\left [{{\partial^2 }\over {\partial t^2}}-(c^2+v^2_A){{\partial^2 }\over {\partial x^2}}\right ] u_y=-\left [f{{\partial}\over {\partial t}}-c^2{{\partial^2 }\over {\partial x \partial y}}\right ] u_x,
\end{equation}
where $v_A=B_x/\sqrt{4\pi\rho}$ is the Alfv\'en speed and $c=\sqrt{gH}$ is the surface gravity speed. Fourier expansion of Equations ~(\ref{eq6}) and (\ref{eq7}) with $exp(-i\omega t +ik_x x)$, leads  to the equation
$$
{{d^2 u_y}\over {d y^2}}+{{2k^4_xc^2v_Av{'}_A}\over {(\omega^2-k^2_xv^2_A) [\omega^2-(c^2+v^2_A)k^2_x] }}{{d u_y}\over {d y}}+\\
$$
\begin{equation}\label{main}
\left [{{\omega^2-k^2_x(c^2+v^2_A)}\over {c^2}}-{{\omega^2f^2}\over {c^2(\omega^2-k^2_xv^2_A)}} -{{k_x\omega f{'}}\over {\omega^2-k^2_xv^2_A}}-{{2k^3_x\omega f v_Av{'}_A}\over {(\omega^2-k^2_xv^2_A)[\omega^2-(c^2+v^2_A)k^2_x] }}\right ] u_y=0,
\end{equation}
where $'$ sign shows differentiation by $y$.

In the local frame at the latitude $\theta_0$, the Coriolis parameter can be expanded as $f=2\Omega\sin{\theta_0}+2\Omega\cos{\theta_0}(\theta-\theta_0)+...$. Retaining only the first order term in the expansion (the so called $\beta$-plane approximation) we get the Coriolis parameter near the equator as
\begin{equation}\label{beta}
f=\beta y,
\end{equation}
where (not to be confused with the plasma $\beta$)
\begin{equation}\label{beta-p}
\beta={{2\Omega}\over {R}}.
\end{equation}

The structure and consequent solutions of Equation~(\ref{main}) depend on the latitudinal structure of the magnetic field.

\section{Equatorial shallow water MHD waves in the solar tachocline with uniform toroidal magnetic field}

Let us suppose that the magnetic field is uniform, i.e. $B_x=B_0=constant$. Then Equation~(\ref{main}) is simplified to
\begin{equation}\label{constant}
{{d^2 u_y}\over {d y^2}}+
\left [{{\omega^2-k^2_x(c^2+v^2_A)}\over {c^2}} - {{k_x\omega \beta}\over {\omega^2-k^2_xv^2_A}}-\mu^2y^2\right ] u_y=0,
\end{equation}
where
\begin{equation}\label{mu}
\mu={{\omega\beta}\over {c\sqrt{\omega^2-k^2_xv^2_A}}}.
\end{equation}
This is the equation of a parabolic cylinder (also known as the equation of the quantum harmonic oscillator), and when
\begin{equation}\label{disp-hom}
{{\omega^2-k^2_x(c^2+v^2_A)}\over {c^2}} - {{k_x\omega \beta}\over {\omega^2-k^2_xv^2_A}}=|\mu|(2n+1)
\end{equation}
then it has bounded solutions \citep{abramowitz}
\begin{equation}\label{solution-hom}
u_y=C\exp\left [-{{|\mu|y^2}\over {2}} \right ]H_n\left (\sqrt{|\mu|}y\right ),
\end{equation}
where $H_n$ is the Hermite polynomial of order $n$ and $C$ is a constant. The solutions are oscillatory inside the interval
\begin{equation}\label{trap-hom}
y<\left |\sqrt{{{2n+1}\over {|{\mu}|}}}\right |
\end{equation}
and exponentially tend to zero outside. 

Equation~(\ref{trap-hom}) shows that the waves are trapped near the equator for small $c=\sqrt{gH}$ and $n$. Therefore, in order to have equatorially trapped shallow water waves one needs a very small surface gravity speed, i.e., a very small reduced gravity, $G$. This means that the waves can be trapped near the equator only in the upper overshoot part of the tachocline. In the lower radiative part of the tachocline the reduced gravity is still large, so the shallow water waves cannot be trapped near the equator, but rather penetrate to higher latitudes (for the same reason the HD shallow water waves considered by \citet{Lou2000} in the solar photosphere were not really concentrated near the equator but were extended up to latitudes $\pm 60$ latitudes). Therefore, in the rest of the paper, we will consider only the upper overshoot layer of the tachocline, where the reduced gravity is very small and hence creates excellent conditions for the trapping of shallow water waves near the equator.

Equation~(\ref{disp-hom}) defines the dispersion relation for equatorial MHD shallow water waves as
\begin{equation}\label{disp}
(\omega^2-k^2_xv^2_A)(\omega^2-k^2_x(c^2+v^2_A))-k_x\beta c^2 \omega=\beta c |\omega| (2n+1) \sqrt{\omega^2-k^2_xv^2_A}.
\end{equation}

\begin{figure}
\plotone{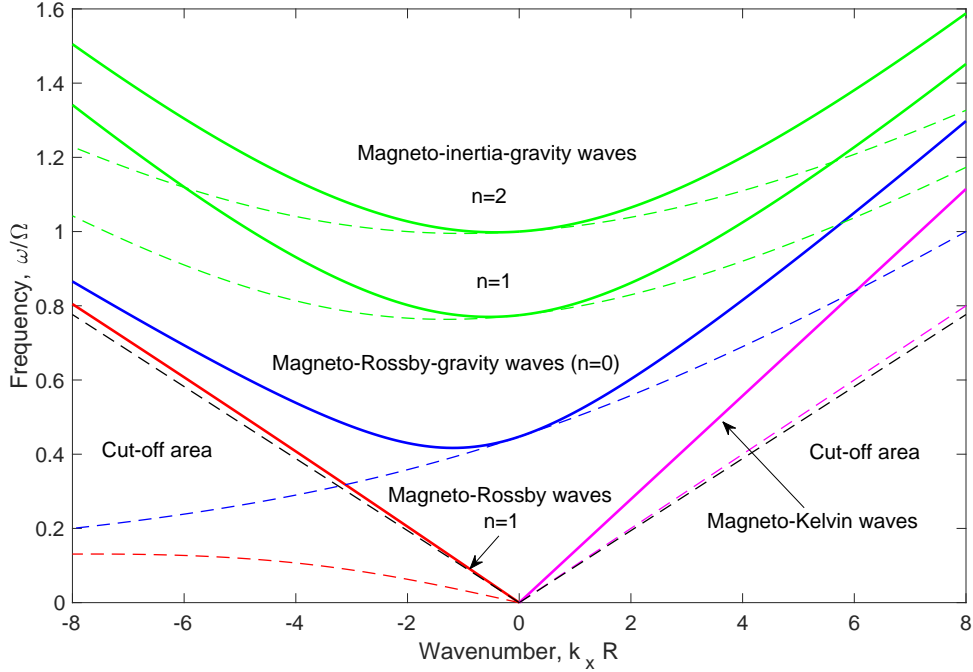}
\caption{Dispersion curves of equatorial shallow water waves in the case of a uniform toroidal magnetic field with strength 20 kG. Reduced gravity is $G=0.01$. The solid green lines correspond to magneto-inertia-gravity waves with n=1, 2. The thick blue solid line corresponds to mixed magneto-Rossby-gravity waves with $n=0$. The solid red line corresponds to magneto-Rossby waves with $n=1$. The solid magenta line corresponds to magneto-Kelvin waves. The dashed green, blue, red, and magenta lines correspond to HD inertia-gravity, mixed Rossby-gravity, Rossby and Kelvin waves respectively. The dashed black lines correspond to pure Alfv\'en wave dispersion curves. \label{fig:fig4}}
\end{figure}

When the magnetic field is zero, then the dispersion relation transforms into the dispersion relation of equatorial HD waves \citep{Matsuno1966}. This is the fourth order algebraic equation and its analytical solution is further complicated owing to the square root on the right hand-side of the equation. The numerical solution also needs certain caution as it may lead to spurious zeroes. The accepted strategy is that one should remove the square root by taking square of both side of the equation and find the solutions of resulted polynomial. This procedure obviously adds four additional solutions, which are not real. Therefore, obtained solutions should be carefully checked and verified if they satisfy the initial equation. After the careful check we obtained the numerical solutions of the dispersion relation in terms of different wave modes which are plotted on Figure 1 for positive frequency. This figure also shows the dispersion curves of HD shallow water waves denoted by dashed lines.

\subsection{Magneto-inertia-gravity and magneto-Rossby waves for $n\geq 1$}

For $n\geq 1$, the solutions include magneto-inertia-gravity and magneto-Rossby waves. The remarkable difference with regard to the HD case is that the uniform toroidal magnetic field creates low frequency cut-off areas in the dispersion relation. The cut-off area is defined by $\omega=\pm k_xv_A$ lines (black dashed lines on Figure 1), which is also clearly seen in Equation~(\ref{disp}) as the right-hand-side term becomes imaginary. This means that the low-frequency Rossby wave solution, which is plotted as the red dashed line in Figure 1 in the HD case, is no longer a solution of the dispersion relation as it is prohibited by the horizontal magnetic field. This can be understood physically as the action of a Lorentz force on the vorticity of the Rossby waves. The strong magnetic field opposes the vortex motion and prohibits the Rossby waves. As the magnetic field strength weakens the low-frequency Rossby waves may still arise, but a magnetic field with strength $> $10 kG blocks their appearance. The solution for the magneto-Rossby waves is just above the Alfv\'en wave solution $\omega=- k_xv_A$. The phase and group speeds of the magneto-Rossby waves are directed eastward and hence the waves propagate opposite to the solar rotation. Magneto-inertia-gravity waves are both westward and eastward propagating (both phase and group speeds are in the same direction) and they have the same dynamics as in the HD case, affected by the magnetic field mostly for higher wave numbers. For $k_x=0$, which actually means pure poleward (or equatorward) propagation, only magneto-inertia-gravity waves remain and their dispersion relation is $\omega=\sqrt{\beta c (2n+1)}$.

\subsubsection{Magneto-inertia-gravity and magneto-Rossby-gravity waves for n=0}

Equation~(\ref{disp}) needs special treatment for the $n=0$ case and can be factorized for  $\omega>0$ as 
\begin{equation}\label{zero-}
 (\sqrt{\omega^2-k^2_xv^2_A}-k_xc)\left [(\omega^2-k^2_xv^2_A)(\sqrt{\omega^2-k^2_xv^2_A}+k_xc)-\beta c \omega \right ]=0.
\end{equation}
The first term of Equation~(\ref{zero-}) is zero when $\omega=\sqrt{c^2+v^2_A}k_x$ for positive $k_x$. However, this solution is spurious as it leads to zero coefficients in front of the velocity perturbations on the left-hand side of Equations~(\ref{eq6}) and (\ref{eq7}).
The second term of Equation~(\ref{zero-}) leads to a real solution for positive and negative $k_x$  (blue solid line in Figure 1). The solution for negative $k_x$ approaches the magneto-Rossby wave curve in the high-wavenumber limit. The phase and group speeds are directed eastward like the magneto-Rossby waves, therefore this is a mixed magneto-Rossby-gravity mode. On the other hand, for positive $k_x$ the solution approaches the magneto-inertia-gravity wave curve in the high-wavenumber limit with westward-directed phase and group speeds. However, there are two remarkable differences between the waves in the MHD (blue solid line) and HD (blue dashed line) cases. First, the magnetic field creates a lower-frequency cut-off area for mixed Rossby-gravity waves with negative wavenumber. Second, the group velocity of HD mixed Rossby-gravity wave is always westward, while the velocity becomes eastward for mixed magneto-Rossby-gravity waves.

\subsection{Magneto-Kelvin waves}

When the poleward components of velocity ($u_y$) and magnetic field  ($b_y$) are zero, then a particular class of solutions magneto-Kelvin waves arises similar to those for HD Kelvin waves. Equations~(\ref{eq1})-(\ref{eq5}) are now rewritten as
\begin{equation}\label{kelvin1}
{{\partial u_x}\over {\partial t}}=-g{{\partial h}\over {\partial x}}+{B_x\over {4\pi\rho}}{{\partial b_x}\over {\partial x}},
\end{equation}
\begin{equation}\label{kelvin2}
{{\partial b_x}\over {\partial t}}=B_x{{\partial u_x}\over {\partial x}},
\end{equation}
\begin{equation}\label{kelvin3}
{{\partial h}\over {\partial t}}+H{{\partial u_x}\over {\partial x}}=0,
\end{equation}
\begin{equation}\label{kelvin4}
fu_x=-g{{\partial h}\over {\partial y}}.
\end{equation}
Fourier expansion of  Equations~(\ref{kelvin1})-(\ref{kelvin3}) with $exp(-i\omega t +ik_x x)$ leads to the dispersion relation
\begin{equation}\label{eq2K1}
(\omega-\sqrt{c^2+v^2_A}k_x)(\omega+\sqrt{c^2+v^2_A}k_x)=0,
\end{equation}
which yields two different modes. The solutions are
\begin{equation}\label{kelvin5}
u_x=u_0\exp\left ({{1\over 2}{{\beta \sqrt{c^2+v^2_A}}\over {c^2}}y^2}\right )
\end{equation}
for the $\omega=-\sqrt{c^2+v^2_A}k_x$ mode and
\begin{equation}\label{kelvin6}
u_x=u_0\exp\left (-{{1\over 2}{{\beta \sqrt{c^2+v^2_A}}\over {c^2}}y^2}\right )
\end{equation}
for the $\omega=\sqrt{c^2+v^2_A}k_x$ mode.
It is seen that the first solution does not satisfythe  boundary condition, therefore it is ruled out. Hence, only the mode with the dispersion relation
\begin{equation}\label{kelvin7}
\omega=\sqrt{c^2+v^2_A}k_x
\end{equation}
remains as the solution for magneto-Kelvin waves (magenta line in Figure 1).

\begin{figure}
\plotone{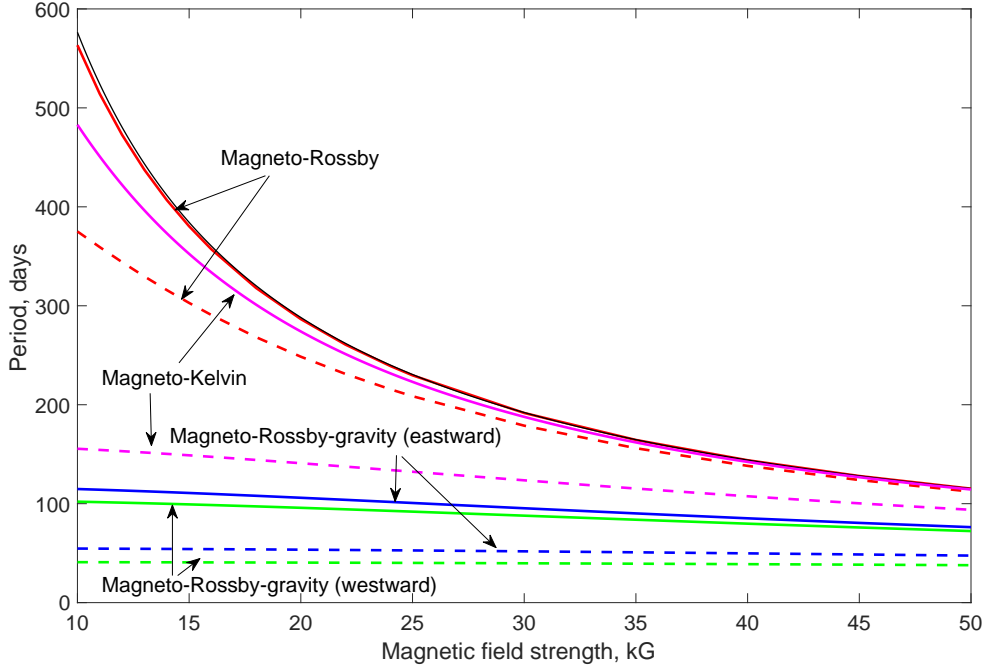}
\caption{Period of equatorial MHD shallow water waves vs. the strength of the uniform magnetic field for the harmonics with $k_x R=1$. The red solid (dashed) line corresponds to $n=1$ magneto-Rossby waves for $G=0.001$ ($G=0.03$). The blue solid (dashed) line corresponds to eastward-propagating $n=0$ magneto-Rossby-gravity waves for $G=0.001$ ($G=0.03$). The green solid (dashed) line corresponds to westward-propagating $n=0$ magneto-Rossby-gravity waves for $G=0.001$ ($G=0.03$). The magenta solid (dashed) line corresponds to westward-propagating $n=-1$ magneto-Kelvin waves for $G=0.001$ ($G=0.03$). The black line corresponds to the Alfv\'en wave solution, $\omega = \pm k_x v_A$. \label{fig:fig5}}
\end{figure}

This solution also arises from the general dispersion relation Equation~(\ref{disp}) if one substitutes $n=-1$. Therefore, this mode was called the $n=-1$ mode by \citet{Matsuno1966}. 

Figure 2 displays the dependence of shallow water wave periods on magnetic field strength. It is seen that magneto-Rossby and magneto-Kelvin waves yield the periods close to the Rieger periodicity for a field strength of $< $ 50 kG. The low-frequency cut-off is also seen (black line) for the strong magnetic field, which is consequently removed for the weaker field.

\section{Equatorial shallow water MHD waves in the solar tachocline with nonuniform toroidal magnetic field}

Now we consider a latitudinally nonuniform toroidal magnetic field. Differential rotation inevitably leads to a nonuniform toroidal field in the case of any poloidal weak magnetic field.

The solar observed latitudinal differential rotation can be expressed as
\begin{equation}
\label{omega} \Omega_d=\Omega (1-s_2 \sin^2\theta-s_4\sin^4\theta),
\end{equation}
where $s_2$ and $s_4$ parameters determined by observations whose values at the solar surface (and in the tachocline) are about 0.13-0.14.

The toroidal component of the induction equation   
 \begin{equation}\label{b-phi} 
 {{\partial B_{\phi}}\over {\partial t}} \sim B_{\theta}{{\partial \Omega_d}\over {\partial \theta}}
\end{equation}
allows us to derive the latitudinal profile of the generated toroidal magnetic field as
\begin{equation}
\label{magnetic} B_{\phi}\sim B_{\theta}(2 \Omega t_0) s_2 \cos \theta \sin \theta,
\end{equation}
where $t_0$ is the time scale of the process. Therefore, the most obvious latitudinal profile of the toroidal field is $\cos \theta \sin \theta$, which has been frequently used for the tachocline, and initiated by \citet{Gilman1997}.

This profile can be approximated near the equator as
\begin{equation}\label{mag}
B_x=B_0{{y}\over {R}}.
\end{equation}
Inserting this expression into Equation~(\ref{main}) and keeping only the terms with $y^2$ we get the equation
\begin{equation}\label{main1}
{{d^2 u_y}\over {d y^2}}+{{2k^4_xc^2v^2_{A0}}\over {R\omega^2(\omega^2-c^2k^2_x)}}{{y}\over R}{{d u_y}\over {d y}}+
\left [{{\omega^2}\over {c^2}}-k^2_x -{{k_x \beta}\over {\omega}} - \left ({{k^2_xv^2_{A0}}\over {R^2c^2}}+{{\beta^2}\over {c^2}}+{{k^3_x\omega \beta v^2_{A0}}\over {\omega^4R^2}}+{{2k^3_x\omega \beta v^2_{A0}}\over {R^2\omega^2(\omega^2-k^2_xc^2)}} \right )y^2\right ] u_y=0.
\end{equation}

By substitution of the expression 
\begin{equation}\label{sub}
u_y={\tilde u_y} \exp \left [{-{{{k^4_xc^2v^2_{A0}}\over {R^2\omega^2(\omega^2-c^2k^2_x)}}}}{{y^2}\over {2}}\right ]
\end{equation}
this equation transforms into the standard form
\begin{equation}\label{final}
{{d^2 {\tilde u_y}}\over {d y^2}}+
\left [{{\omega^2}\over {c^2}}-k^2_x -{{k_x \beta}\over {\omega}} - {{k^4_xc^2v^2_{A0}}\over {R^2\omega^2(\omega^2-c^2k^2_x)}}-{\tilde \mu}^2y^2\right ]{\tilde u_y}=0,
\end{equation}
where
\begin{equation}\label{mu1}
{\tilde \mu}=\sqrt{{{k^2_xv^2_{A0}}\over {R^2c^2}}+{{\beta^2}\over {c^2}}+{{k^3_x\omega \beta v^2_{A0}}\over {\omega^4R^2}}+{{2k^3_x\omega \beta v^2_{A0}}\over {R^2\omega^2(\omega^2-k^2_xc^2)}}+{{k^8_xc^4v^4_{A0}}\over {R^4\omega^4(\omega^2-c^2k^2_x)^2}}}.
\end{equation}

When 
\begin{equation}\label{bounded}
{{\omega^2}\over {c^2}}-k^2_x -{{k_x \beta}\over {\omega}} - {{k^4_xc^2v^2_{A0}}\over {R^2\omega^2(\omega^2-c^2k^2_x)}}=|{\tilde \mu}|(2n+1)
\end{equation}
then Equation~(\ref{final}) has the bounded solution
\begin{equation}\label{solution-magneto}
{\tilde u_y}=C\exp\left [-{{|{\tilde \mu}| y^2}\over {2}} \right ]H_n\left (\sqrt{|{\tilde \mu}|}y\right ),
\end{equation}
where $H_n$ is a Hermite polynomial of order $n$ and $C$ is a constant. The solutions are oscillatory inside the interval
\begin{equation}\label{trap-magneto}
y<\left |\sqrt{{{2n+1}\over {|{\tilde \mu}|}}}\right |
\end{equation}
and exponentially tend to zero outside.  Equation~(\ref{bounded}) defines the dispersion relation
\begin{equation}\label{disp-inhom}
{{\omega^2}\over {c^2}}-k^2_x -{{k_x \beta}\over {\omega}} - {{k^4_xc^2v^2_{A0}}\over {R^2\omega^2(\omega^2-c^2k^2_x)}}=(2n+1)\sqrt{{{k^2_xv^2_{A0}}\over {R^2c^2}}+{{\beta^2}\over {c^2}}+{{k^3_x\omega \beta v^2_{A0}}\over {\omega^4R^2}}+{{2k^3_x\omega \beta v^2_{A0}}\over {R^2\omega^2(\omega^2-k^2_xc^2)}}+{{k^8_xc^4v^4_{A0}}\over {R^4\omega^4(\omega^2-c^2k^2_x)^2}}}.
\end{equation}
This is a six order polynomial equation with $\omega$, therefore its exact analytical solution is very complicated, also owing to the square root on the right-hand side. The numerical solution also should be performed with caution (see the previous section). However, before proceeding, it is possible to find analytical solutions to some approximation. One can observe that $\omega=\pm k_x c$ is the singular point. It is seen from Equations~(\ref{eq6}) and (\ref{eq7}) that this solution at the equator yields zero coefficients in front of velocity perturbations on the left-hand side (actually, the singular point corresponds to magneto-Kelvin waves as we will see later). Therefore, we can look to approximate solutions for $ |\omega| \ll  |k_x c|$ and $|\omega| \gg |k_x c|$.

\subsection{Magneto-inertia-gravity waves ($|\omega| \gg | k_x c |$)}

In this approximation, one can get the dispersion relation
\begin{equation}\label{disp-big}
{{\omega^3}}-\left (k^2_x c^2+ (2n+1)c\sqrt{{{k^2_xv^2_{A0}}\over {R^2}}+{{\beta^2}}}\right )\omega-{{k_x \beta c^2}} =0,
\end{equation}
which is similar to that of HD shallow water waves, and for the weak magnetic field approximation it completely transforms into the HD inertia-gravity wave solution.

\subsection{Magneto-Rossby waves ($|\omega| \ll | k_x c| $)}

The other limit of the lower-frequency waves  ($|\omega| \ll | k_x c| $) leads to the dispersion relation
\begin{equation}\label{disp-rossby}
k_x c \omega  -  {{k^2_xcv^2_{A0}}\over {\beta R^2}}=-(2n+1)\sqrt{\omega^4 - {{2k_xv^2_{A0}}\over { \beta R^2}} \omega^3+ {{k^3_xc^2  v^2_{A0}}\over {\beta R^2}}  \omega +{{k^4_xc^2v^4_{A0}}\over {\beta^2 R^4 }}}.
\end{equation}
This is a second-order polynomial with $\omega$ and it corresponds to magneto-Rossby waves. The magnetic field splits the equatorial HD Rossby waves into fast and slow magneto-Rossby waves (see also \citet{zaqarashvili2007,zaqarashvili2009}). The dispersion relation for the high-frequency fast modes can be further simplified to
\begin{equation}\label{disp-rossby-fast}
{{\omega_{+}}} \approx - {{k_xc}\over {2n+1}}.
\end{equation}
On the other hand, the dispersion relation for the low-frequency slow modes can be simplified to 
\begin{equation}\label{disp-rossby-slow}
 \omega_{-} \approx - {{(2n+1)^2-1}\over {(2n+1)^2+2}} {{k_xv^2_{A0}}\over {\beta R^2}}.
\end{equation}
The fast magneto-Rossby waves are similar to HD Rossby waves, while the slow magneto-Rossby waves are a new type which arise owing to both Coriolis and Lorentz forces.

\begin{figure}
\plotone{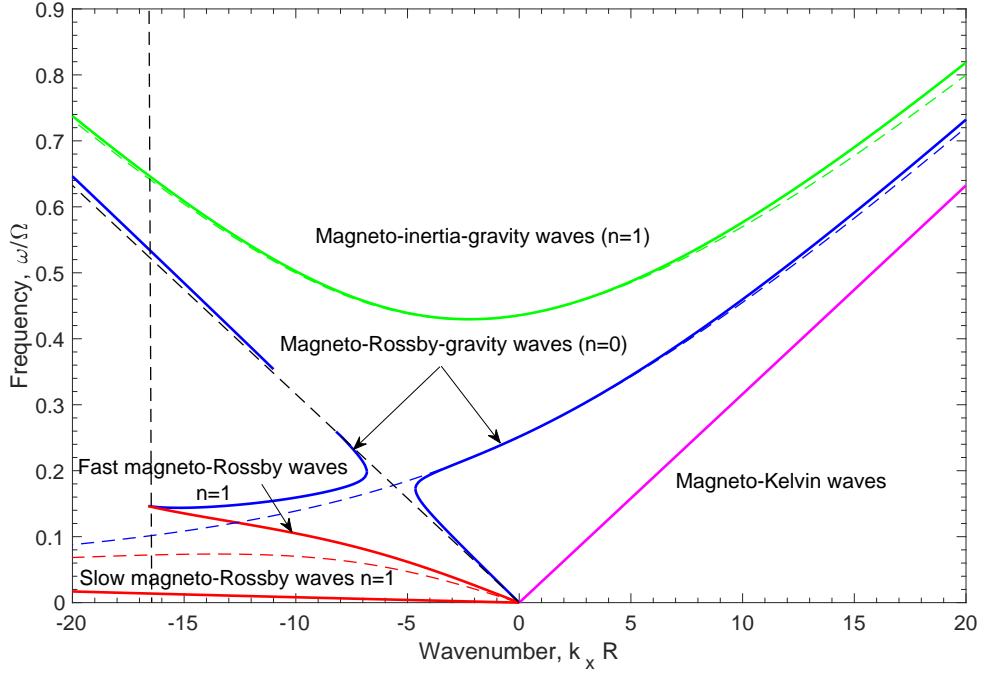}
\caption{Dispersion curves of equatorial shallow water waves in the case of a nonuniform magnetic field ($B_0=10$ kG). The green solid (dashed) line corresponds to $n=1$ magneto-inertia-gravity (HD inertia-gravity) waves. The blue solid (dashed) lines correspond to $n=0$ magneto-Rossby-gravity (HD Rossby-gravity) waves. The red solid (dashed) lines correspond to $n=1$ magneto-Rossby (HD Rossby) waves. The solid magenta line corresponds to magneto-Kelvin waves at the equator. The vertical black dashed line indicates the cut-off wavenumber for fast magneto-Rossby and slow magneto-Rossby-gravity waves. The inclined black dashed line corresponds to the curve $\omega=-k_x c$. Reduced gravity is $G=0.001$. \label{fig:f5}}
\end{figure}

\subsection{Magneto-Kelvin waves}

In the case of the magnetic field profile (Equation~\ref{mag}), Equations~(\ref{kelvin1})-(\ref{kelvin4}) are rewritten after Fourier transformation as
\begin{equation}\label{kelvin11}
\omega u_x=g k_x h -{B_{0}\over {4\pi\rho}}{y\over {R}} k_x b_x,
\end{equation}
\begin{equation}\label{kelvin21}
\omega  b_x =-B_{0}{y\over {R}} k_x u_x,
\end{equation}
\begin{equation}\label{kelvin31}
\omega h -H k_x u_x =0,
\end{equation}
\begin{equation}\label{kelvin41}
\beta y u_x=-g{{\partial h}\over {\partial y}}.
\end{equation}

Eqs.~(\ref{kelvin11})-(\ref{kelvin31}) lead to the dispersion relation
\begin{equation}\label{eq2KM}
\left (\omega-\sqrt{c^2+v^2_{A0}{y^2\over {R^2}}}k_x \right )  \left (\omega+\sqrt{c^2+v^2_{A0}{y^2\over {R^2}}}k_x \right )=0,
\end{equation}
which yields two different modes. The solutions for each mode can be easily found from Equations~(\ref{kelvin31}) and (\ref{kelvin41}) for ${y^2\over {R^2}} \ll 1$
\begin{equation}\label{kelvin51}
u_x=u_0 \exp \left (-{1\over 2}\left ({\beta \over c} - {v^2_{A0}\over {c^2 R^2}}\right ) y^2 \right )
\end{equation}
for the $\omega=\sqrt{c^2+v^2_{A0}{y^2\over {R^2}}}k_x$ mode and
\begin{equation}\label{kelvin61}
u_x=u_0 \exp \left ({1\over 2}\left ({\beta \over c} - {v^2_{A0}\over {c^2 R^2}}\right ) y^2\right )
\end{equation}
for the $\omega=-\sqrt{c^2+v^2_{A0}{y^2\over {R^2}}}k_x$  mode. When ${\beta \over c} > {v^2_{A0}\over {c^2 R^2}}$, then the first solution satisfies the boundary conditions, hence it is the solution in the weak magnetic field limit. However, for the very strong magnetic field with
${\beta \over c} < {v^2_{A0}\over {c^2 R^2}}$  the second solution satisfies the boundary conditions. For a magnetic field strength of  $<$  100 kG the mode with the dispersion relation
\begin{equation}\label{kelvinM}
\omega=\sqrt{c^2+v^2_{A0}{y^2\over {R^2}}}k_x
\end{equation}
remains the solution for magneto-Kelvin waves. It is seen that the mode has different frequencies at different $y$ layers. At the equator, its dispersion relation is
\begin{equation}\label{kelvinM1}
\omega=k_x c
\end{equation}
similar to the HD case. This is expected since the magnetic field vanishes at $y=0$.

\begin{figure}
\plotone{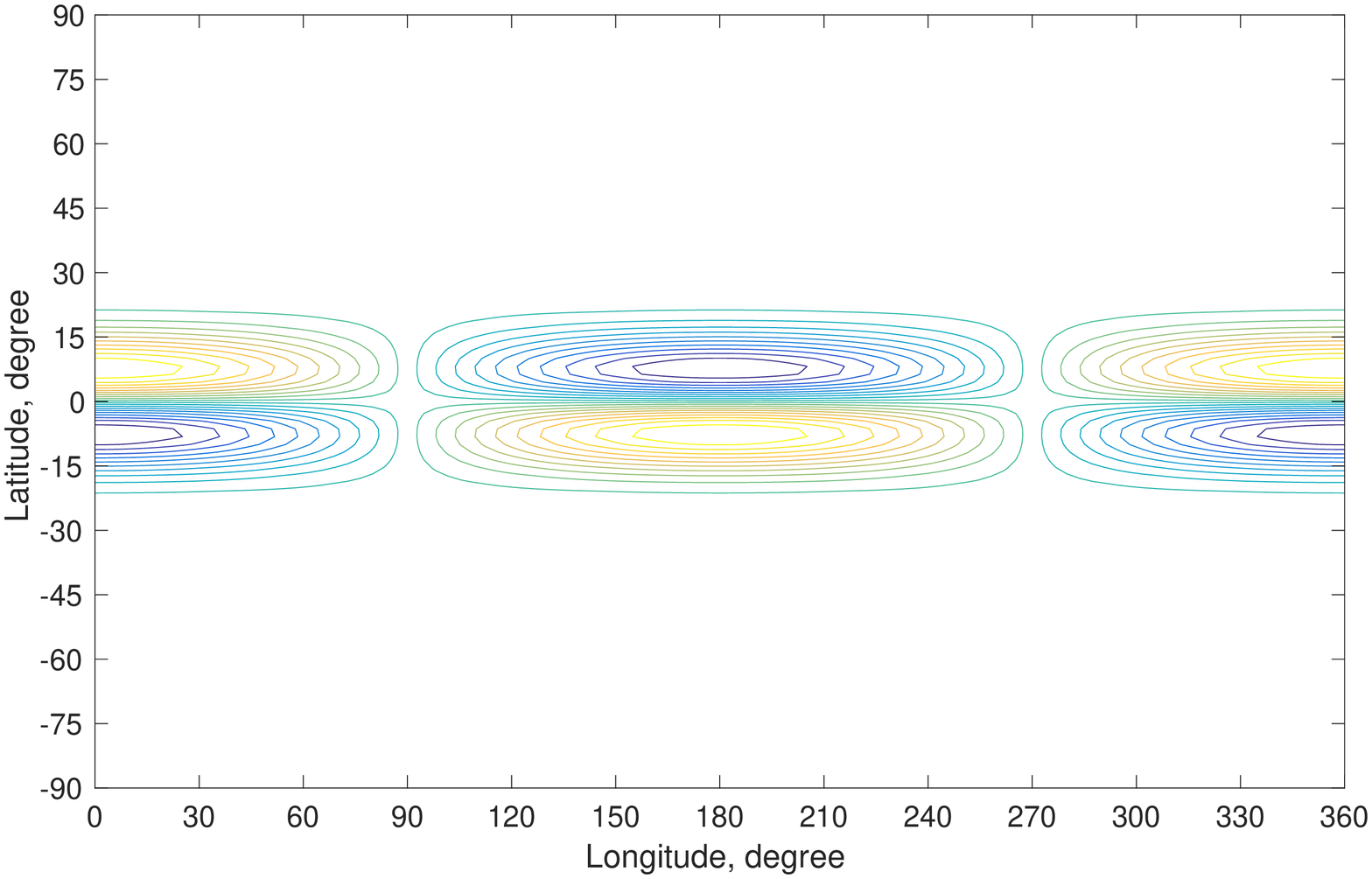}
\plotone{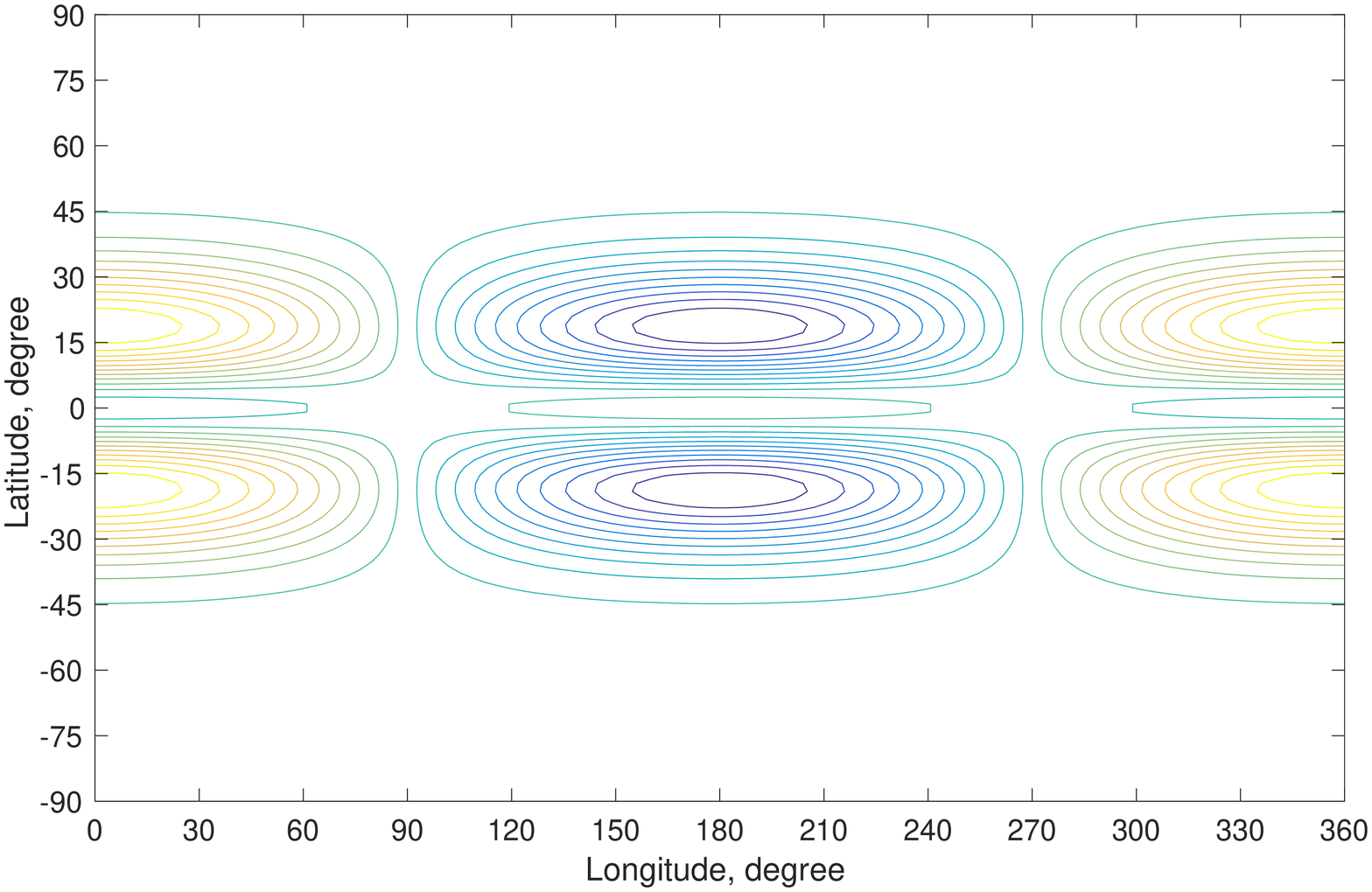}
\caption{Contour plot of the toroidal component of the perturbed magnetic field, $b_x$ (normalised by $B_0$) in fast magneto-Rossby waves with $n=1$ (upper panel) and $n=2$ (lower panel). Here $B_0=10$ kG, $G=0.001$ and $k_x R=1$. \label{fig:f71}}
\end{figure}

The numerical solution of the dispersion relation, Equation~(\ref{disp-inhom}), should again be performed with caution. We take the square of both sides and solve the resulting 12th-order polynomial numerically. Spurious solutions are carefully determined and removed from the consideration. The resulting real solutions are plotted in Figure 3, which shows the very different timescales of magneto-inertia-gravity and magneto-Rossby waves (note that the corresponding HD wave modes are plotted as the dashed lines). The mixed magneto-Rossby-gravity ($n$=0) and magneto-Kelvin ($n$=-1) waves have intermediate timescales. The magnetic field has almost no influence on the frequency of the inertia-gravity and Kelvin waves, but it leads to the splitting of the HD Rossby waves (red dashed line) into fast and slow magneto-Rossby waves (red solid lines). The westward-propagating magneto-Rossby-gravity waves are not affected by the magnetic field. On the other hand, eastward-propagating magneto-Rossby-gravity waves (with negative wavenumber) are split into fast and slow modes (blue solid lines) by the magnetic field for small and large wavelengths. The magnetic field also leads to the appearance of several cut-off areas. The first appears for eastward magneto-Rossby-gravity waves in the wavelength interval of $7> |k_xR| >4$ for the fixed parameters of magnetic field and reduced gravity. The second area arises for the fast magneto-Rossby-gravity waves in the wavelength interval of $11> | k_xR | >8$, where the mode crosses the dispersion curve $\omega=-k_x c$ (inclined black dashed line), which is a spurious solution of the system. The third, probably most important, cut-off area appears for the fast magneto-Rossby and slow magneto-Rossby-gravity waves for large wavenumbers $| k_xR | >16.5$ (vertical black dashed line). The phase speed of the fast and slow magneto-Rossby waves is directed to the east i.e., the waves propagate in the opposite direction to rotation. Their group speeds are also directed eastward. This is in contrast to the HD Rossby waves, which have westward-directed group speed for large wavelengths \citep{gill82}. Here the magnetic field blocks the westward group speed of the Rossby waves. The fast and slow magneto-Rossby-gravity waves ($n=0$) propagate eastward, while the $n=0$ magneto-inertia-gravity waves propagate westward. For small wavelengths, the group speeds of the fast and slow magneto-Rossby-gravity waves are directed westward and eastward, respectively. For large wavelengths, however, the group speed of the fast magneto-Rossby-gravity waves is directed eastward and that of the slow waves is directed westward. The magneto-inertia-gravity waves propagate east- and westward, while the magneto-Kelvin waves propagate only westward. The group speeds of both types of waves have the direction of their phase speeds.

\begin{figure}
\includegraphics[width = 3.7in]{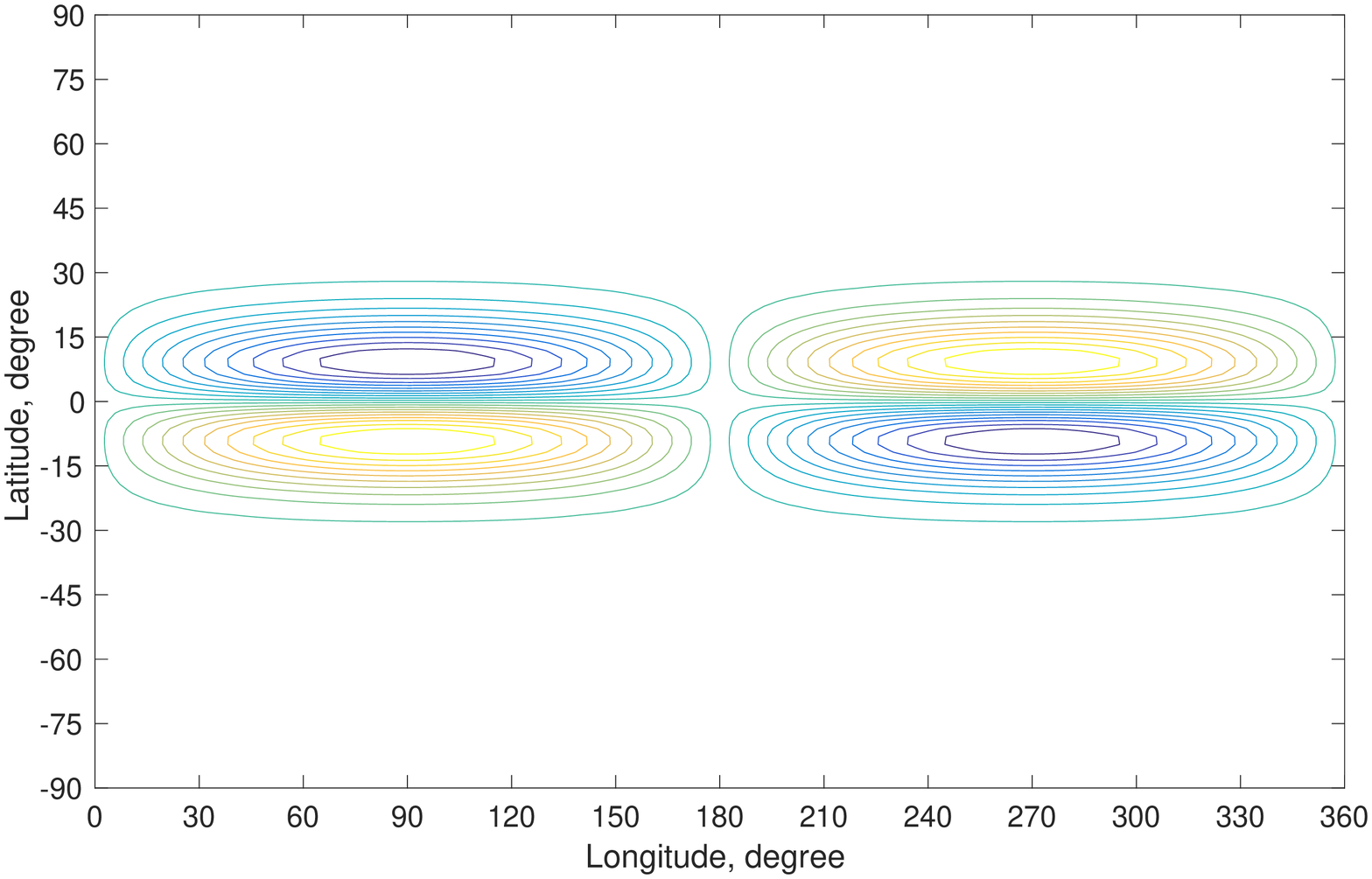}
\includegraphics[width = 3.7in]{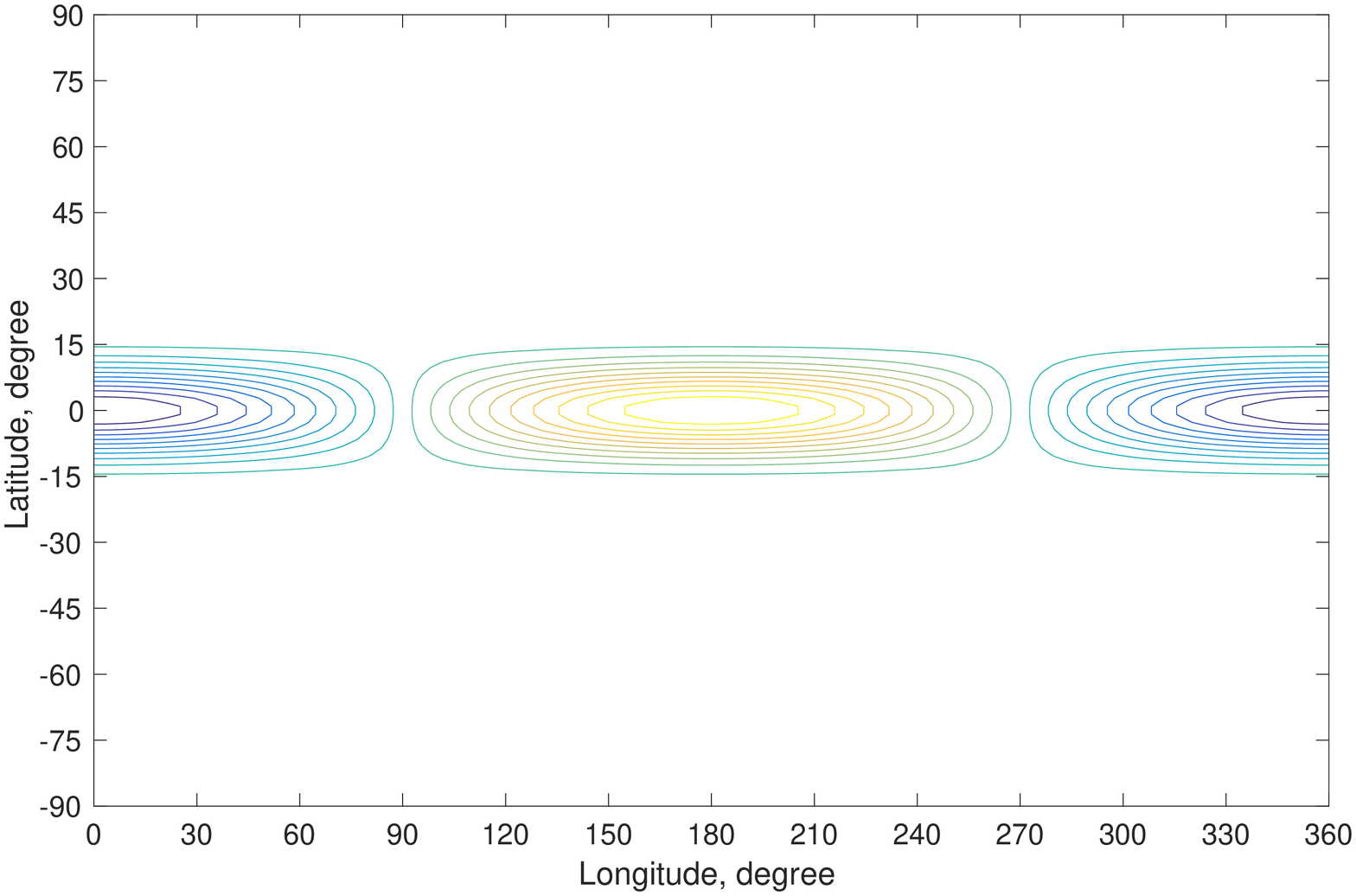}
\includegraphics[width = 3.7in]{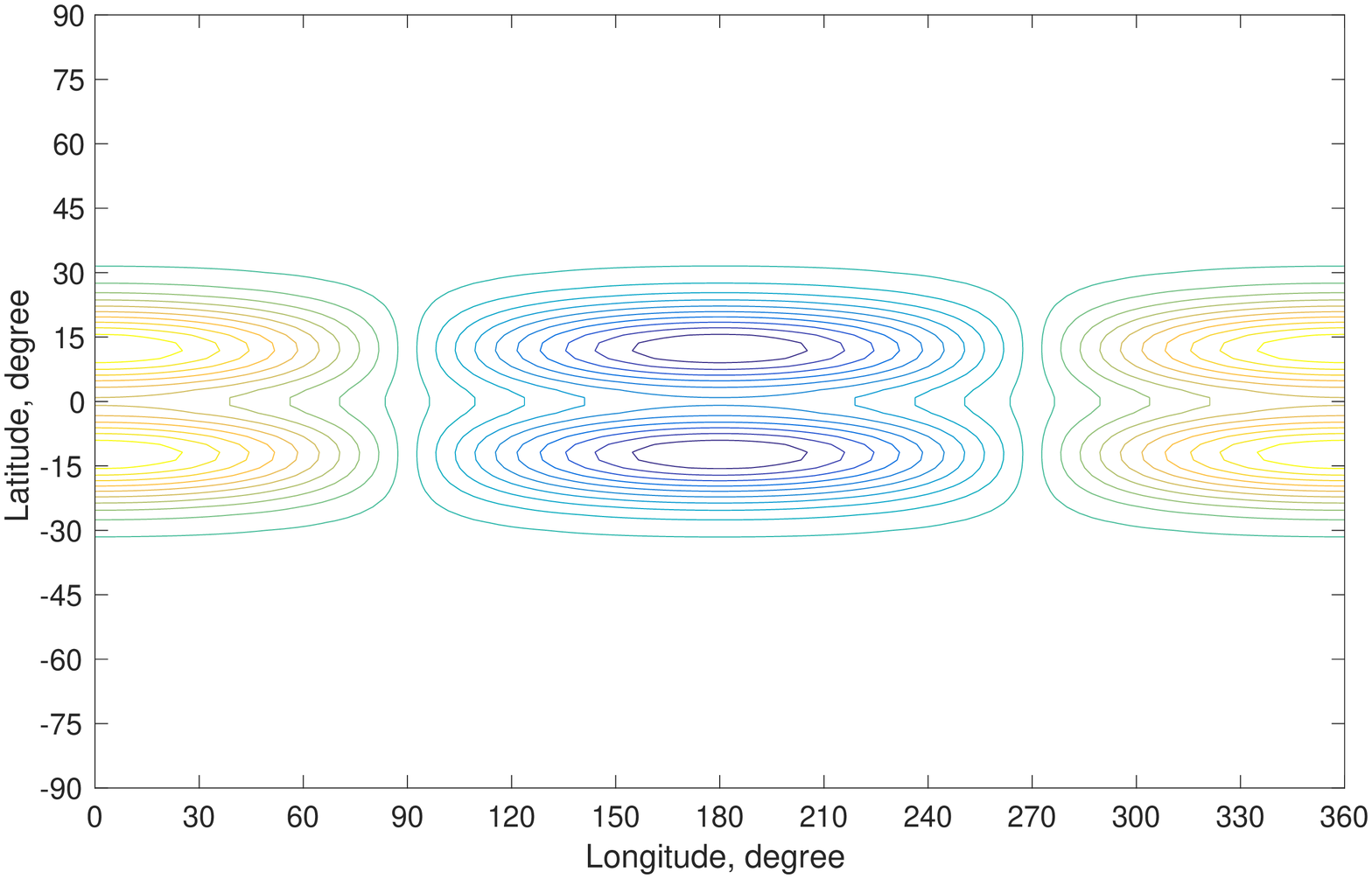}
\includegraphics[width = 3.7in]{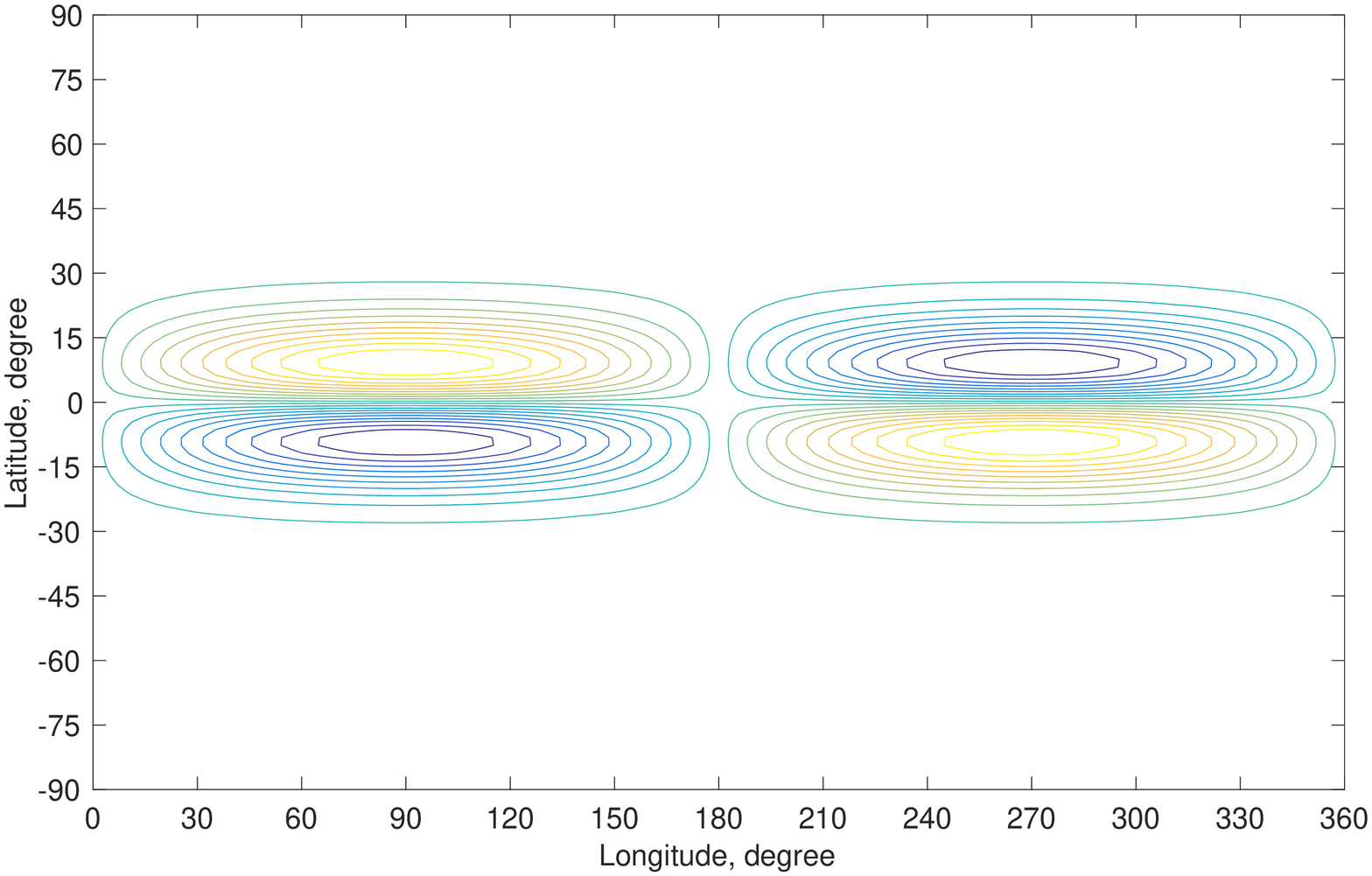}
\caption{Contour plots of poloidal velocity $u_y$ (normalized by $R \Omega$, upper left panel), toroidal velocity $u_x$ (normalized by $R \Omega$, upper right panel), layer thickness  $h$ (normalized by $H$, lower left pane),  and poloidal magnetic field $b_y$ (normalized by $B_0$, lower right panel)  in fast magneto-Rossby waves with $n=1$. Here $B_0=10$ kG, $G=0.001$ and $k_x R=1$. \label{fig:f61}}
\end{figure}

The solutions of all shallow water waves are confined near the equator (for detailed spatial structures of HD equatorial waves see, e.g., \citet{bouchut2005}). Figure 4 shows the contour plots of the $n=1,2$ harmonics of the fast magneto-Rossby waves when the unperturbed toroidal magnetic field ($B_0$) is 10 kG, the normalized reduced gravity ($G$) is 0.001 and the normalized toroidal wave number ($k_x R$) is 1. The $n=1$ harmonic is sandwiched between latitudes $\pm 20^0$, has one zero (at the equator) as expected, and consequently opposite signs in the northern and southern hemispheres. The $n=2$ harmonic is sandwiched between latitudes $\pm 40^0$ and has two zeros near latitudes $\pm 5^0$.

Figure 5 shows the contour plots of all other variables in the fast magneto-Rossby waves (with $n=1$) with the same parameters as in Figure 4. The perturbation of the poleward magnetic field $b_y$ has the same spatial structure (lower right panel in Figure 5) as the poleward velocity (upper left panel in Figure 5), but with a180$^0$ shift in the toroidal direction. This is clearly expected from Equation~(\ref{eq4}). The solutions for the perturbations of layer thickness and toroidal velocity have the same sign in the two hemispheres, but they obviously change sign in the toroidal direction and have 180$^0$ phase shift. 

The splitting of HD Rossby waves into fast and slow magneto-Rossby waves is displayed in Figure 6. When the wavelength reaches the toroidal extent of the equator (i.e., $k_xR =1$) then the period of the fast magneto-Rossby waves with $n=1$ is about 7 yr for a field strength of 10 kG and normalized reduced gravity of 0.001. For the wavelength $k_xR =0.5$ the period is about 11 yr, exactly matching the solar cycle length. At the same time, the period of the slow magnetic Rossby waves for $k_xR =1$ reaches a value of 95 years, which is in the range of the Gleissberg cycle \citep{Gleissberg1939}. The period of the magneto-Kelvin waves is about 2 yr, hence it is in the range of annual oscillations. Therefore, equatorial MHD shallow water waves cover almost all observed timescales of solar long-term activity variations in the expected conditions of the overshoot layer.

\begin{figure}
\plotone{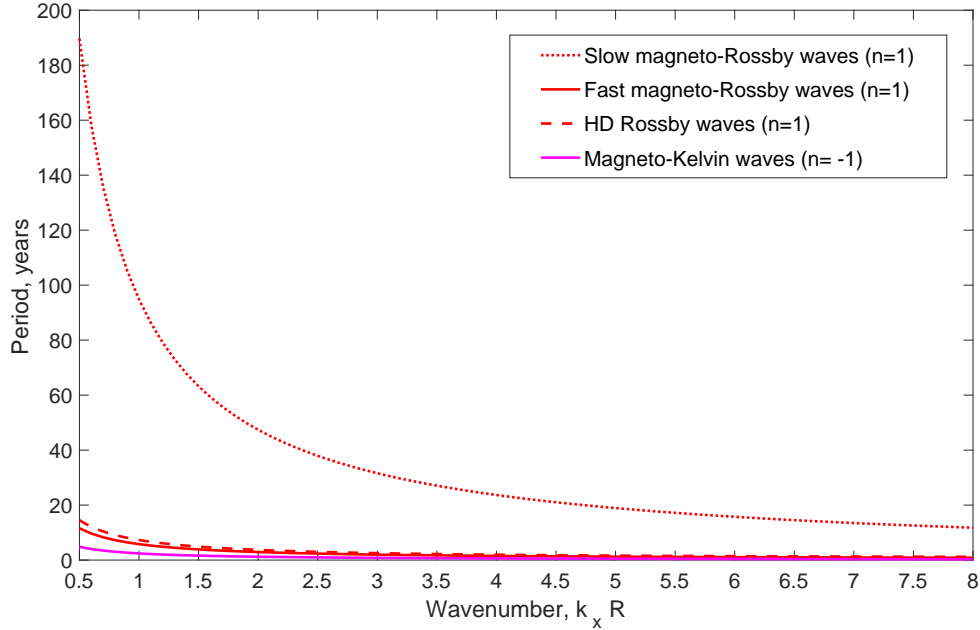}
\caption{Period of equatorial magneto-Rossby and magneto-Kelvin waves vs. toroidal wavelength. The thick red line corresponds to fast magneto-Rossby waves. The red dotted line corresponds to slow magneto-Rossby waves. The red dashed line corresponds to HD Rossby waves. The magenta line corresponds to magneto-Kelvin waves at the equator. Here $B_0=10$ kG and $G=0.001$. \label{fig:f6}}
\end{figure}

\section{Discussion}

Solar activity undergoes variations over different timescales from tens/hundreds of days to tens/hundreds of years. The most pronounced variation occurs with a period of $\sim$ 11 yr, which is usually explained by dynamo theory concerning differential rotation and convection. Besides the 11 yr cycles, several other periodicities are seen in different activity indices. Long-term modulation of cycle amplitude with a periodicity of 100 yr (and more) is seen in sunspot numbers and cosmogenic radionuclides. On the other hand, shorter-period variations are detected as annual (with periods of 1?2 yr) and mid-range  ($< $ 200 days) oscillations. The physical mechanisms for long- and short-term variations have not generally been determined. On the other hand, the Rossby wave scenario in the solar interior may capture the essential physics of the variation. 

Rossby (or planetary) waves owe their excitation to the conservation of absolute vorticity in a rotating fluid. They lead to the formation of cyclones/anticyclones in higher latitudes of the Earth, which actually govern weather over Europe and the USA. The waves are also trapped near the equator owing to the minimum of the Coriolis parameter (so-called equatorially trapped or equatorial waves), which might lead to the observed long-period oscillations in oceans. Therefore, Rossby waves generally determine the large-scale dynamics of the Earth's atmosphere and oceans. 

\begin{figure}
\plotone{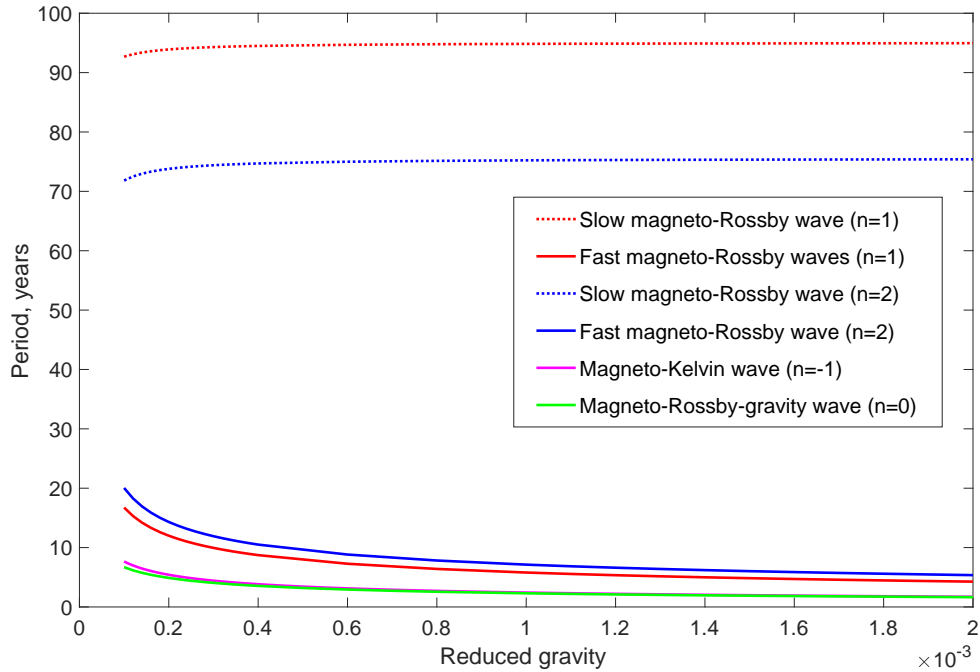}
\caption{Period of equatorial magneto-Rossby, slow magneto-Rossby-gravity, and magneto-Kelvin waves vs. reduced gravity. The red (blue) solid line corresponds to $n=1$ ($n=2$) fast magneto-Rossby waves. The red (blue) dotted line corresponds to  $n=1$ ($n=2$) slow magneto-Rossby waves. The green line corresponds to slow magneto-Rossby-gravity wave (n=0). The magenta line corresponds to magneto-Kelvin waves at the equator. Here $B_0=10$ kG and $k_x R=1$. \label{fig:f7}}
\end{figure}

On the other hand, Rossby waves might be important for the large-scale dynamics of solar/stellar interiors, where their properties are modified by large-scale magnetic fields. Rossby waves have been studied in the presence of a toroidal magnetic field and differential rotation in the solar tachocline.  \citet{Lou2000} studied the dispersion relations for various equatorially trapped HD shallow water waves in the solar photosphere, but the dynamics of equatorial magneto-Rossby waves has remained unexplored until now. In this paper, we studied the dynamics of equatorial shallow water waves in the solar tachocline for different latitudinal profiles of the toroidal magnetic field. The initial MHD shallow water equations lead to the equation of a parabolic cylinder near the equator, which has a solution in terms of Hermite polynomials satisfying the boundary conditions of equatorial confinement. This consideration allowed us to obtain the dispersion relations of various MHD shallow water waves for different magnetic field profiles.

First, the equatorial HD shallow water waves previously described by \citet{Lou2000} in the photosphere lead to essentially different periodicities in the upper overshoot tachocline (see the dashed lines in Figures 1 and 3). The difference is related to the reduced gravity, which comes from the sub-adiabatic temperature gradient in the tachocline \citep{Gilman2000,Dikpati2001}. The typical reduced gravity in the overshoot layer tends to the confinement of shallow water waves near the equator (the waves are extended up to 60$^0$ in the case of normal gravity) and leads to a significant increase of Rossby and Rossby-gravity wave periods. The normal gravity considered by \citet{Lou2000} in the photosphere yields a period of equatorial Rossby waves in the range of one hundred days, while the typical reduced gravity ($G \sim$ 0.001) in the overshoot region yields a period of $\sim$ 7 yr for the Rossby waves, hence approaching the timescale of solar cycles in the case of $G \sim$ 0.0004. On the other hand, the inertia-gravity waves show the Rieger-type periodicity of one hundred days in the case of the reduced gravity.

Second (Section 3), the latitudinally uniform toroidal magnetic field has a significant influence on the dispersion relations of equatorial shallow water waves. The main result of the influence is that the magnetic field creates a low-frequency cut-off region forbidding the appearance of long-period Rossby and Rossby-gravity waves. A magnetic field of strength $> $10 kG blocks the appearance of low-frequency Rossby waves owing to the action of the Lorentz force on their vorticity. The magnetic field may also stabilize the baroclinic instability, as recently studied by \citet{Zeitlin2013}. This point needs future detailed investigation also in the solar context. However, a latitudinally uniform toroidal magnetic field is probably not a good approximation for the solar dynamo layer as the observed latitudinal differential rotation will inevitably lead to a nonuniform toroidal component.

\begin{figure}
\plotone{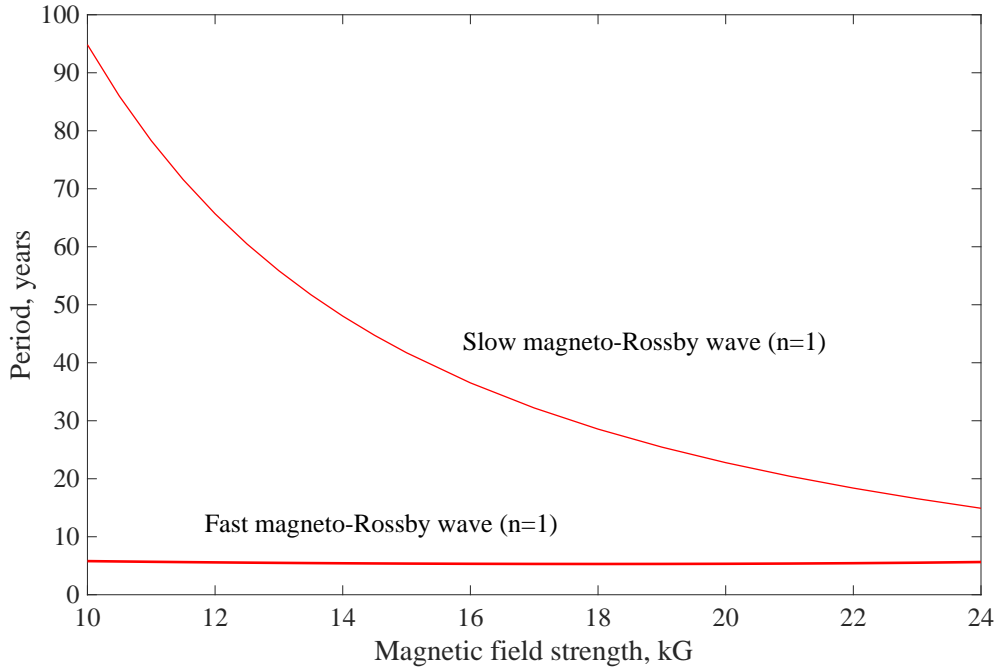}
\caption{Period of equatorial fast (thick solid line) and slow (thin solid line) magneto-Rossby waves vs. toroidal magnetic field strength. Here $G=0.001$ and $k_x R=1$. \label{fig:f9}}
\end{figure}

Third (Section 4), the non-uniform toroidal magnetic field, which resembles the solar magnetic field, affects the dispersion relations of shallow water waves in the tachocline conditions. We chose a latitudinal profile of  $\sim \cos \theta \sin \theta$ for the toroidal magnetic field. This profile is actually obtained by the action of the observed differential rotation on a poloidal field component. It can be approximated as $\sim y$ near the equator in the Cartesian coordinate system. A sixth-order polynomial dispersion equation (Equation~\ref{disp-inhom}) is obtained, which describes all equatorial shallow water waves such as magneto-Rossby, magneto-Rossby-gravity, magneto-inertia-gravity, and magneto-Kelvin waves. Simple analytical expressions and numerical solutions of the general dispersion equation are derived, which show very interesting timescales for all waves. The magnetic field splits the ordinary HD Rossby and Rossby-gravity waves into fast and slow modes, while the magneto-inertia-gravity waves are almost identical to their HD counterparts (see Figure 3). The magnetic field creates several cut-off areas for magneto-Rossby and magneto-Rossby-gravity waves. The most important cut-off area appears for large wavenumbers of fast magneto-Rossby and slow magneto-Rossby-gravity waves. The cut-off wavelength is 190 Mm for a magnetic field strength of 10 kG and reduced gravity of $G=$ 0.001  (see Figure 3). It would be also interesting to study how the non-uniform magnetic field will influence the baroclinic instability. We will briefly discuss each mode and the corresponding observed timescale of solar activity variations.

\subsection{Equatorial fast magneto-Rossby waves and Schwabe cycle}

Solar activity undergoes 11 yr oscillations known as Schwabe cycles \citep{Schwabe1844}. These are generally interpreted in terms of dynamo models, but the interpretation still faces important problems  \citep{Charbonneau2005}. An equatorial fast magneto-Rossby wave has a similar timescale under certain conditions of the overshoot layer. Figure 7 shows the dependence of magneto-Rossby, magneto-Rossby-gravity, and magneto-Kelvin waves on the normalized value of reduced gravity,  $G$, for a field strength of 10 kG. We can see that $G=0.0003-0.0004$ yields the period of fast magneto-Rossby waves with $n=1$ similar to the Schwabe cycles i.e. $\sim$ 11 yr. The period of fast magneto-Rossby waves does not significantly depend on the strength of the toroidal magnetic field (see Figure 8), therefore it is generally determined by the value of reduced gravity, which in turn is due to the sub-adiabatic temperature gradient.

The latitudinal extent of the solutions depends on the normalized reduced gravity $G$ (lower latitudes with smaller $G$), therefore the observed equatorward drift of sunspots can be obtained if either the temperature gradient or tachocline thickness is changing through the cycle. This cannot be obtained by the simple linear analysis considered in this paper but future nonlinear consideration my reveal the observed dynamics of sunspots.

\subsection{Equatorial slow magneto-Rossby mode and Gleissberg cycle}

Long-term records of sunspot numbers have revealed long-period modulation of Schwabe cycles over 80-100 yr \citep{Gleissberg1939,Hathaway2010}. This modulation was explained in terms of slow magneto-Rossby waves excited in the lower part of the tachocline \citep{zaqarashvili2015}. Figures 7 and 8 show that equatorial slow magneto-Rossby waves indeed lead to the observed modulation timescale. In contrast to the fast magneto-Rossby waves, the slow waves significantly depend on toroidal magnetic field strength. For a field strength of 10 kG, the period of $n=1$ slow magneto-Rossby waves tends to 90-100 yr, which is in excellent coincidence with the Gleissberg cycle. The superposition of fast and slow magneto-Rossby modes may lead to the observed long-term modulation of activity cycles.

\begin{figure}
\plotone{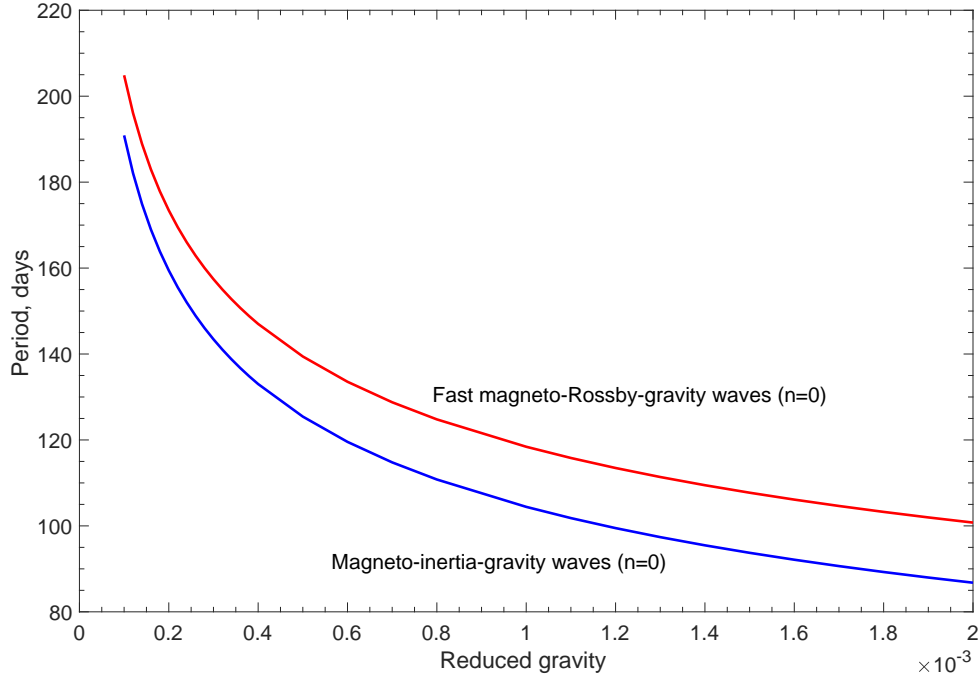}
\caption{Period of fast magneto-Rossby-gravity (red) and magneto-inertia-gravity (blue) waves with $n=0$ vs. reduced gravity. Here $B_0=10$ kG and $k_x R=1$. \label{fig:f8}}
\end{figure}

\subsection{Equatorial magneto-Kelvin and slow magneto-Rossby-gravity waves versus annual/quasi-biennial oscillations}

Oscillations with period $\sim$ 2 yr (quasi-biennial oscillations) are found in almost all indices of solar activity \citep{sakurai81,vecchio10}. The oscillations were suggested to be connected with magneto-Rossby wave instability in the solar tachocline \citep{zaqarashvili2010b}. Helioseismology also revealed oscillations of solar tachocline velocity with period 1.3 yr \citep{Howe2000}. Recent observations of coronal bright points based on STEREO and SDO data also showed annual oscillations \citep{McIntosh2015}, which were interpreted by Rossby waves in the solar interior  \citep{McIntosh2017}. However, our results (see Figures 6 and 7) show that equatorial magneto-Kelvin and slow magneto-Rossby-gravity ($n$=0) waves have timescales of 1-2 yr, which coincide with observed annual oscillations. The magneto-Kelvin and magneto-Rossby-gravity waves do not significantly depend on the unperturbed magnetic field strength. Therefore, $G=0.001$ yields the period of magneto-Kelvin and slow magneto-Poincare-Rossby waves as 2 yr for wavenumber $k_x R=1$ (or $m=1$) and 1 yr for $k_x R=2$ (or $m=2$). Again, there is a very nice coincidence with observed periodicity.

\subsection{Equatorial fast magneto-Rossby-gravity and magneto-inertia-gravity waves versus Rieger-type periodicity}

A short periodicity between 152 and 158 days was discovered in $\gamma$- and X-ray flares during solar cycle 21 by \citet{Rieger1984}. Since the discovery, numerous works observed this periodicity in different indices of solar activity \citep{carball90,Oliver1998,Gurgenashvili2016,Gurgenashvili2017}. It was explained by magneto-Rossby wave instability in the solar tachocline \citep{zaqarashvili2010a, Dikpati2017}. However, our results indicate that in the framework of equatorial waves the periodicity is connected with magneto-inertia-gravity and/or fast magneto-Rossby-gravity waves rather than with magneto-Rossby waves. Figure 9 shows that $n=0$ eastward-propagating fast magneto-Rossby-gravity wave and westward-propagating magneto-inertia-gravity waves have a Rieger-type periodicity for $G=0.0003-0.0004$. 

\section{Conclusion}

We studied the linear dynamics of shallow water waves in the solar tachocline in the presence of a toroidal unperturbed magnetic field. It is shown that the reduced gravity owing to the sub-adiabatic temperature gradient in the upper overshoot layer of the tachocline leads to confinement of the waves near the equator. The dispersion relations of the equatorial waves are obtained for latitudinally uniform and non-uniform toroidal magnetic field profiles. The dispersion relations are solved analytically and numerically, describing the dynamics of various shallow water waves. A reasonable value of the temperature gradient in the overshoot region leads to an increase in the period of equatorial Rossby waves up to the timescale of solar cycles. A latitudinally uniform magnetic field creates a low-frequency cut-off region forbidding the occurrence of low-frequency Rossby and Rossby-gravity waves. A latitudinally non-uniform toroidal magnetic field (generated by the observed latitudinal differential rotation from a weak poloidal component) leads to both shorter and longer timescales of shallow water waves. The magnetic field splits ordinary HD Rossby waves into fast and slow magneto-Rossby modes and eastward-propagating $n=0$ magneto-Rossby-gravity waves into fast and slow modes. It also blocks the appearance of fast magneto-Rossby and slow magneto-Rossby-gravity waves for large wavenumbers. On the other hand, the inertia-gravity and Kelvin waves are not significantly affected by the magnetic field. With a reasonable value of reduced gravity in the overshoot region, the harmonics with $k_x R \approx 0.5-1$ of fast magneto-Rossby waves have the period of Schwabe cycles i.e., 11 yr, which suggests the possible role of equatorial Rossby waves in the generation of solar cycles. The solutions are concentrated between latitudes $\pm 20^0-40^0$ which coincide with those of sunspot appearance. In the same case, the period of equatorial slow magneto-Rossby waves is similar to that of the Gleissberg cycle i.e.,  $\sim$ 100 yr. On the other hand, the periods of magneto-Kelvin (and slow magneto-Rossby-gravity) and magneto-inertia-gravity (and fast magneto-Rossby-gravity) waves correspond to observed annual/quasi-biennial and Rieger-type oscillations, respectively. Therefore, all modes of equatorial magneto shallow water waves reflect almost all observed periodicities in solar activity. Future analytical/numerical/observational study is required to resolve this long-standing solar activity problem.

\acknowledgments

This work was supported by the Austrian Science Fund (FWF, projects P26181-N27 and P30695-N27) and by Georgian Shota Rustaveli National Science Foundation project 217146. The very useful advice of an unknown referee, which led to significant improvement of the paper, is warmly acknowledged. This paper resulted from discussions at the workshop of ISSI (International Space Science Institute) team (ID 389) "Rossby Waves in Astrophysics" organized in Bern (Switzerland).




\begin{thebibliography}{}

\bibitem[Abramowitz \& Stegun(1964)]{abramowitz}Abramowitz, M., \& Stegun, I.A. 1964, {\it Handbook of Mathematical Functions} (Washington, D.C.: National Bureau of Standards)
\bibitem[Andersson(1999)]{Andersson1999}Andersson, N., Kokkotas, K., and Schutz, B. F., 1999, \apj, 510, 846
\bibitem[Bonomo and Lanza(2012)]{Bonomo2012}Bonomo, A.S. and Lanza, A.F., 2012, \aa, 547, A37
\bibitem[Bouchut et al.(2005)]{bouchut2005}Bouchut, F., Le Sommer, J. and Zeitlin, V., 2005,  Chaos, 15, 013503
\bibitem[Carbonell \& Ballester(1990)]{carball90} Carbonell, M., \& Ballester, J.L. 1990, \aap, 238, 377
\bibitem[Charbonneau(2005)]{Charbonneau2005}Charbonneau, P. 2005, LRSP, 2, 2 
\bibitem[Dikpati and Gilman(1999)]{Dikpati1999}Dikpati, M., \&  Gilman, P. A. 1999, \apj, 512, 417
\bibitem[Dikpati and Gilman(2001)]{Dikpati2001}Dikpati, M., \&  Gilman, P. A. 1999, \apj, 551, 536
\bibitem[Dikpati et al.(2003)]{Dikpati2003}Dikpati, M., Gilman, P. A., \& Rempel, M. 2003, \apj, 596, 680
\bibitem[Dikpati et al.(2006)]{Dikpati2006}Dikpati, M., Gilman, P. A. and MacGregor, K.B. 2006, \apj, 638, 564
\bibitem[Dikpati et al.(2007)]{Dikpati2007}Dikpati, M., Gilman, P. A., de Toma, G. and Ghosh, S. S., 2007, Solar Phys., 245, 1
\bibitem[Dikpati et al.(2017)]{Dikpati2017}Dikpati, M., Cally, P. S., McIntosh, S. W., and Heifetz, E., 2017, Nature Rep., 7, 14750
%\bibitem[Gigolashvili et al.(1995)]{gigolashvili95}Gigolashvili, M. Sh., Japaridze, D. R., Pataraya, A. D. \& Zaqarashvili, T.V., 1995, \solphys, 156, 221
\bibitem[Gill(1982)]{gill82}Gill, A. E. 1982, Atmosphere-Ocean Dynamics, San Diego: Academic Press
\bibitem[Gilman and Fox(1997)]{Gilman1997}Gilman, P. A. and Fox, O. A., 1997, \apj, 484, 439
\bibitem[Gilman(2000)]{Gilman2000}Gilman, P. A. 2000, \apj, 544, L79
\bibitem[Gleissberg(1939)]{Gleissberg1939}Gleissberg, M. N., 1939, Observatory, 62, 158
\bibitem[Gurgenashvili et al.(2016)]{Gurgenashvili2016}Gurgenashvili, E., Zaqarashvili, T. V., Kukhianidze, V., et al., 2016, ApJ, 826, 55
\bibitem[Gurgenashvili et al.(2017)]{Gurgenashvili2017}Gurgenashvili, E., Zaqarashvili, T. V., Kukhianidze, V., et al., 2017, ApJ, 845, 137
\bibitem[Hathaway(2010)]{Hathaway2010}Hathaway, D. H., 2010, Living Rev. Solar Phys. 7, 1 URL: http://solarphysics.livingreviews.org/Articles/lrsp-2010-1
\bibitem[Heifetz et al.(2015)]{Heifetz2015}Heifetz, E., Mak, J., Nycander, J. and Umurhan, O. M., 2015, J. Fluid Mech., 767, 199
\bibitem[Heng and Spitkovsky(2009)]{Heng2009}Heng, K. and Spitkovsky, A., 2009, \apj, 703, 1819
\bibitem[Heng and Workman(2014)]{Heng2014}Heng, K. and Workman, J.,, 2014, \apjs, 213, 27
\bibitem[Howe et al.(2000)]{Howe2000}Howe, R., Christensen-Dalsgaard, J., Hill, F., et al.,  2000, Science, 287, 2456
\bibitem[Klimachkov and Petrosyan(2017)]{Klimachkov2017}Klimachkov, D. A. and Petrosyan, A. S., 2017, Phys. Let. A, 381, 106
\bibitem[Lanza et al.(2009)]{Lanza2009}Lanza, A. F., Pagano, I., Leto, G. et al., 2009, \aa, 493, 193
\bibitem[Longuet-Higgins(1968)]{Longuet-Higgins1968}Longuet-Higgins, M. S., 1968, Proc. R. Soc. London. A., 262, 511
\bibitem[Lovelace and Hohlfeld(1978)]{Lovelace1978}Lovelace, R. V. E. and Hohlfeld, R. G., 1978, \apj, 221, 51
\bibitem[Lovelace et al.(1999)]{Lovelace1999}Lovelace, R. V. E., Li, H., Colgate, S. A. and Nelson, A. F., 1999, \apj, 513, 805
\bibitem[Lou(1987)]{Lou1987}Lou, Y. Q. 1987,  \apj, 322, 862
\bibitem[Lou(2000)]{Lou2000}Lou, Y.Q.  2000, \apj, 540, 1102
\bibitem[Lou(2001)]{Lou2001}Lou, Y. Q. 2001, \apjl, 563, L147
\bibitem[Lou et al.(2003)]{Lou2003}Lou, Y. Q., Wang, Y. M., Fan, Z., et al. 2003, MNRAS, 345, 809
\bibitem[Matsuno(1966)]{Matsuno1966}Matsuno, T.,1966, J. Meteorol. Soc. Japan, 44, 25
\bibitem[McIntosh et al.(2015)]{McIntosh2015}McIntosh, S. W., Leamon, R.J., Krista, L.D. et al., 2015, Nature Communications, 6, 6491
\bibitem[McIntosh et al.(2017)]{McIntosh2017}McIntosh, S. W., Cramer W. J., Marcano M. P., and Leamon, R. J., 2017, Nature Astronomy, 1, 0086
\bibitem[Oliver et al.(1998)]{Oliver1998}Oliver, R., Ballester, J. L., \&  Boudin, F.  1998, Nature, 394, 552
\bibitem[Petviashvili(1980)]{Petviashvili1980}Petviashvili, V. I., 1980, JETP Letters, 32, 619
\bibitem[Rieger et al.(1984)]{Rieger1984}Rieger, E.,  Share, G. H., Forrest, D. J., Kanbach, G., Reppin, C., et al. 1984, Nature, 312, 623
\bibitem[Rossby(1939)]{Rossby1939}Rossby, C.-G., 1939, J. Marine Research, 2, 38
\bibitem[Sakurai(1981)]{sakurai81}Sakurai, K., 1981, \solphys, 74, 35
\bibitem[Schecter et al.(2001)]{Schecter2001}Schecter, D. A., Boyd, J. F. and Gilman, P. A., 2001, \apj, 551, L185
\bibitem[Schwabe(1844)]{Schwabe1844}Schwabe, H., 1844, Astron. Nachr. 21, 233
\bibitem[Spiegel and Zahn(1992)]{Spiegel1992}Spiegel, E. A. and Zahn, J.-P., 1992, \aa, 265, 106
\bibitem[Umurhan(2010)]{Umurhan2010}Umurhan, O. M., 2010, \aa, 521, A25
\bibitem[Vecchio et al.(2010)]{vecchio10}Vecchio, A., Laurenza, M., Carbone, V. \& Storini, M., 2010, \apj, 709, L1
\bibitem[Zeitlin(2013)]{Zeitlin2013}Zeitlin, V., 2013, Nonlinear Processes in Geophysics, 20, 893
\bibitem[Zaqarashvili et al.(2007)]{zaqarashvili2007}Zaqarashvili, T. V., Oliver, R., Ballester, J. L., \& Shergelashvili, B. M. 2007, A\&A, 470, 815
\bibitem[Zaqarashvili et al.(2009)]{zaqarashvili2009}Zaqarashvili, T. V., Oliver, R., \& Ballester, J. L. 2009, \apjl, 691, L41
\bibitem[Zaqarashvili et al.(2010a)]{zaqarashvili2010a}Zaqarashvili, T. V., Carbonell, M., Oliver, R., \& Ballester, J. L. 2010a, \apj, 709, 749
\bibitem[Zaqarashvili et al.(2010b)]{zaqarashvili2010b}Zaqarashvili, T. V., Carbonell, M., Oliver, R., \& Ballester, J. L. 2010b, \apjl, 724, L95
%\bibitem[Zaqarashvili et al.(2011)]{zaqarashvili2011}Zaqarashvili, T. V., Oliver, R., Ballester, J. L., Carbonell, M., Khodachenko, M. L. et al., 2011, A\&A, 532, A139
\bibitem[Zaqarashvili et al.(2015)]{zaqarashvili2015}Zaqarashvili, T. V., Oliver, R., Hanslmeier, A., Carbonell, M., Ballester, J. L., et al. 2015, \apjl, 805, L14


\end{thebibliography}
\end{document}